\documentclass[aps,twocolumn,prb,amsmath,amssymb,superscriptaddress]{revtex4-1}
\usepackage{mathrsfs}
\usepackage{multirow}
\usepackage{graphicx,epsfig}

\def \beq {\begin{equation}}
\def \edq {\end{equation}}
\def \bes {\begin{subequations}}
\def \eds {\end{subequations}}
\def \beqn {\begin{equation*}}
\def \edqn {\end{equation*}}

\def \dag {\dagger}

\def \up {\uparrow}
\def \down {\downarrow}

\def \calj {{\cal{J}}}

\begin{document}
\title{Nonlinear spin-thermoelectric transport in two-dimensional topological insulators}
\author{Sun-Yong Hwang}
\affiliation{Institut de F\'{\i}sica Interdisciplin\`aria i Sistemes Complexos
IFISC (CSIC-UIB), E-07122 Palma de Mallorca, Spain}
\affiliation{Pohang University of Science and Technology (POSTECH), Pohang 790-784, Korea}
\author{Rosa L\'opez}
\affiliation{Institut de F\'{\i}sica Interdisciplin\`aria i Sistemes Complexos
IFISC (CSIC-UIB), E-07122 Palma de Mallorca, Spain}
\affiliation{Kavli Institute for Theoretical Physics, University of California, Santa Barbara, California 93106-4030, USA}
\author{Minchul Lee}
\affiliation{Department of Applied Physics, College of Applied Science, Kyung Hee University, Yongin 446-701, Korea}
\author{David S\'anchez}
\affiliation{Institut de F\'{\i}sica Interdisciplin\`aria i Sistemes Complexos
IFISC (CSIC-UIB), E-07122 Palma de Mallorca, Spain}
\affiliation{Kavli Institute for Theoretical Physics, University of California, Santa Barbara, California 93106-4030, USA}

\begin{abstract}
We consider spin-polarized transport in a quantum spin Hall antidot system
coupled to normal leads. Due to the helical nature of the conducting edge states, the screening potential
at the dot region becomes spin dependent without external magnetic fields nor ferromagnetic contacts.
Therefore, the electric current due to voltage or temperature differences becomes spin polarized,
its degree of polarization being tuned with the dot level position or the base temperature.
This spin-filter effect arises in the nonlinear transport regime only and has a purely interaction origin.
Likewise, we find a spin polarization of the heat current which is asymmetric with respect to the bias
direction. Interestingly, our results show that a pure spin current can be generated by thermoelectric means:
when a temperature gradient is applied, the created thermovoltage (Seebeck effect) induces a spin-polarized current
for vanishingly small charge current. An analogous effect can be observed for the heat transport:
a pure spin heat flows in response to a voltage shift even if the thermal current is zero.
\end{abstract}
\maketitle

\section{Introduction}
Two-dimensional topological insulators support gapless current-carrying edge states
characterized by opposite propagation direction for opposite spins.\cite{kan05,ber06}
The conduction of these helical states is protected against disorder since backscattering is forbidden
by time-reversal symmetry.\cite{kan05b,she05,xu06} Therefore, a quantum Hall effect arises
with a two-terminal conductance given by $2e^2/h$, equivalently to the quantum
Hall conductance for filling factor $2$. The difference is that in the quantum {\em spin}
Hall effect the external magnetic field is absent and the edge states arise from
a topologically nontrivial phase in samples with strong spin-orbit coupling.
Experimentally, the quantum spin Hall effect has been confirmed in HgTe/CdTe
heterostructures,\cite{kon07,rot09} showing the spin polarization of the conducting states.\cite{bru12}
In InAs/GaSb quantum wells, quantized transport due to helical states has been observed
even in the presence of external magnetic fields\cite{kne11} and disorder.\cite{kne13}

An exciting consequence of the spatial separation between pairs of helical states
is the emergence of spin filtering effects.\cite{dol11,kru11,cit11,rom12,suk12,dol13,guo14}
However, the spin current in a two-terminal quantum spin Hall bar is zero due to the constrained geometry.
Therefore, backscattering centers are to be implemented to preferably deflect
electrons with a given spin direction. A feasible possibility is the application
of local potentials to form quantum antidots. More generally, the 
presence of constrictions in two-dimensional topological insulators
have been proposed to give rise to coherent oscillations,\cite{chu09}
transformations between ordinary and topological regimes,\cite{tka11}
peaks of noise correlations,\cite{edg13}
metal-to-insulator quantum phase transitions,\cite{cha13}
nonequilibrium fluctuation relations,\cite{lop12}
braiding of Majorana fermions,\cite{mi13}
competition between Fabry-P\'erot and Mach-Zehnder processes,\cite{riz13}
control of edge magnetization,\cite{tim12}
and detection of Kondo clouds.\cite{pos13}
Interestingly, K\"onig {\it et al.} have experimentally
demonstrated\cite{kon13} the local manipulation of helical states with
back-gate electrodes.
\begin{figure}[t]
\centering
\includegraphics[width=0.45\textwidth, clip]{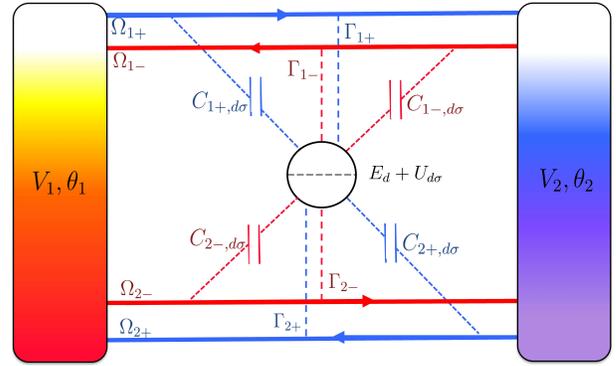}
\caption{(Color online) Schematics of our setup.
A quantum spin Hall bar with a single-level antidot at the center is attached to two terminals, where both voltage bias and temperature gradient are applied. Interactions are described using capacitance coefficients $C_{is,d\sigma}$,
where $i=1,2$ labels the edges, $s=\pm$ is the helicity, $d$ stands for dot,
and $\sigma=\uparrow,\downarrow$ is the electronic spin.
Couplings between the helical edge states and the dot are denoted with $\Gamma_{is}$. 
}\label{fig:1}
\end{figure}

Our aim here is to show that spin-polarized currents can be generated
in quantum spin Hall antidot systems \textit{using thermal gradients only}.
In fact, we demonstrate below that pure spin currents and pure spin heat flows
can be produced by thermoelectric means (Seebeck and Peltier effects).
These effects are relevant because many topological insulators show
excellent thermoelectric properties.\cite{muc13} For instance,
porous three-dimensional topological insulators display large thermoelectric
figures of merit\cite{tre11} and similar properties have been associated
to edge conduction channels\cite{tak10} and nanowires.\cite{goo14}
Moreover, spin Nernst signals can provide spectroscopic information
in quantum spin Hall devices.\cite{rot12}
Here, we consider a simple setup: a two-dimensional topological insulator
connected to two electronic reservoirs, see Fig.~\ref{fig:1}.
The central antidot allows scattering between helical states in different edges,
these transitions preserving the spins of the carriers. Therefore, in the linear
regime of transport and for normal conductors the spin current is zero.
However, in the nonlinear regime the screening potential in the dot region
becomes spin dependent since, quite generally, the dot level will be asymmetrically
coupled to the edge states. As a consequence, the nonlinear current
will be spin polarized. This makes the nonlinear regime of quantum thermoelectric
transport quite unique and interesting to explore, as recently emphasized
in Refs.~\onlinecite{san13,whi13,mea13,lop13,her13,hwa13,mat13,bed13,fah13,dut13,whi14}.

Heat currents can also become spin polarized,
and we find a spin Peltier effect\cite{gra06,flip12} in addition to a spin Seebeck effect.\cite{Uchida,jaw10,sla10}
Rectification effects are more visible in the heat flow,\cite{seg05,cha06,ruo11} which results in
strongly asymmetric spin polarizations. We stress that the spin-filter effects discussed here exist
regardless of couplings to ferromagnetic contacts or external magnetic (Zeeman) fields
(cf. Refs.~\onlinecite{ted71, mes94, bre11, jan12,Vera13}),
and are thereby of purely spintronic\cite{fab07} (or spin caloritronic)\cite{bau10}
character. Furthermore, the spin polarization for both charge and heat currents
can be controlled in our system by adjusting the antidot resonant level
or changing the background temperature.

	The paper is organized as follows.
	In Sec.~\ref{sec:Scr}, we describe our model based on scattering theory to determine the generalized
	transmission probability that depends on the screening potential.
	Intriguingly, the potential response in the antidot region is spin-dependent even though the contacts are normal leads [Eqs.~\eqref{CPu} and \eqref{CPz}], giving rise to spin-polarized electronic and heat currents [Eqs.~\eqref{eq:IsNor} and \eqref{eq:JsNor}], with the asymmetric tunneling described by the parameter $\eta$.
	The transport coefficients are calculated in 
	Sec.~\ref{sec:weakly} using an expansion
	around the equilibrium point. We analytically show that the leading-order rectification terms of the currents with respect to voltage and thermal biases show spin-dependent screening effects, in contrast to the linear coefficients. These results are
	 central to our work. Section~\ref{Numerical}  presents numerical results that are valid beyond the Sommerfeld and the weakly nonlinear approximations when both voltage and thermal biases applied to the sample are strong. We also discuss the possibility
	 of generating pure spin currents from the combination of Seebeck effect and helical propagation in the nonlinear
	 regime of transport. Finally, our conclusions are contained in Sec.~\ref{sec:Con}.

\section{Theoretical model}\label{sec:Scr}
	We consider a quantum spin Hall (QSH) bar attached to two terminals $\alpha=1,2$, where each terminal is driven by the electrical voltage bias $eV_{\alpha}=\mu_{\alpha}-E_{F}$ ($\mu_{\alpha}$ is the electrochemical potential and $E_{F}$ is the common Fermi energy) and also by the temperature shift $\theta_{\alpha}=T_{\alpha}-T$ ($T_{\alpha}$ and $T$ are the lead and the background temperature, respectively), see Fig.~\ref{fig:1}.
	An antidot is formed inside the QSH bar. It can connect upper and lower gapless helical edge states.
	Scattering off the dot is described with the matrix $s_{\alpha\beta}=s_{\alpha\beta}(E, eU)$, which is generally a function of the carrier energy $E$ and the electrostatic potential $U$ inside the system.\cite{but93,chr96}
	The potential $U_{\sigma}=U(\vec{r},\{V_{\gamma}\},\{\theta_{\gamma}\},\sigma)$ is, in turn, a function of the position $\vec{r}$, the set of driving fields $\{V_{\gamma}\}$ and $\{\theta_{\gamma}\}$,\cite{san13,mea13,lop13} and the spin index $\sigma=\up,\down$.
	The $\sigma$-dependence of $U_{\sigma}$ becomes crucial in our QSH system due to the underlying helicity, i.e., the spin-channel separation of charge carriers according to their motion.
	As a matter of fact, the different response of screening potential through the antidot with respect to each spin-component is the working principle for our observed spin-polarized electric and heat currents since these fluxes are determined by
	the spin-dependent potential response.
	
	More specifically, the charge and heat currents at lead $\alpha$ carried by spin-component $\sigma$ are respectively given by\cite{but90}
\begin{align}	
&I_{\alpha}^{\sigma}=\frac{e}{h}\sum_{\beta}\int dE A_{\alpha\beta}^{\sigma}
(E,eU)f_{\beta}(E),\label{I_sigma}\\
&{\cal J}_{\alpha}^{\sigma}=\frac{1}{h}\sum_{\beta}\int dE(E-\mu_{\alpha}) A_{\alpha\beta}^{\sigma}(E,eU)f_{\beta}(E),\label{J_sigma}
\end{align}
where $A_{\alpha\beta}^{\sigma}=\text{Tr}[\delta_{\alpha\beta}-s_{\alpha\beta}^{\dag}s_{\alpha\beta}]$ and $f_{\beta}(E)=(1+\exp[(E-\mu_{\beta})/k_{B}T_{\beta}])^{-1}$ is the Fermi-Dirac distribution function in the reservoir $\beta=1,2$.
	Note here that we have generalized the expressions for charge and heat currents into their spin-resolved form, for which we separate $2A_{\alpha\beta}$ in the usual current expressions\cite{lop13} $I_{\alpha}=(2e/h)\sum_{\beta}\int dE A_{\alpha\beta}(E,eU)f_{\beta}(E)$ and ${\cal J}_{\alpha}=(2/h)\sum_{\beta}\int dE(E-\mu_{\alpha}) A_{\alpha\beta}(E,eU)f_{\beta}(E)$ into $A_{\alpha\beta}^{\up}=A_{\alpha\beta}(U_{\up})$ and $A_{\alpha\beta}^{\down}=A_{\alpha\beta}(U_{\down})$ in order to explicitly incorporate the spin-dependent screening effect.
	
	Due to current conservation for respective $\sigma$ and neglecting spin-flip scattering,\cite{Ste14} one has $\sum_{\alpha}I_{\alpha}^{\sigma}=0$ and $\sum_{\alpha}({\cal J}_{\alpha}^{\sigma}+I_{\alpha}^{\sigma}V_{\alpha})=0$, and one can define the direction of spin-resolved currents: $I_{\sigma}\equiv I_{1}^{\sigma}=-I_{2}^{\sigma}$ and ${\cal J}_{\sigma}\equiv {\cal J}_{1}^{\sigma}=-{\cal J}_{2}^{\sigma}-I_{\sigma}(V_{1}-V_{2})$.
	With this convention, we define the spin-polarized currents
\begin{align}
I_{s}&= I_{\up}-I_{\down} \\
{\cal J}_{s}&= {\cal J}_{\up}-{\cal J}_{\down}
\end{align}
along with the total fluxes $I_{c}\equiv I_{\up}+I_{\down}$ and ${\cal J}_{c}\equiv {\cal J}_{\up}+{\cal J}_{\down}$
(charge and heat, respectively).
	
	The screening potential $U=\sum_{\sigma}U_\sigma$
	is sensitive to variations of the external voltage or temperature biases. Since our theory is based
	on an expansion around the equilibrium point, it suffices to expand the potential
	up to linear order in the driving fields,\cite{san13,mea13,lop13}
\begin{equation}\label{eq:U}
U=U_{\text{eq}}+\sum_{\alpha,\sigma}u_{\alpha\sigma}V_{\alpha}+\sum_{\alpha,\sigma}z_{\alpha\sigma}\theta_{\alpha},
\end{equation}
where $u_{\alpha\sigma}=(\partial U_{\sigma}/\partial V_{\alpha})_{\text{eq}}$ and $z_{\alpha\sigma}=(\partial U_{\sigma}/\partial\theta_{\alpha})_{\text{eq}}$ are spin-dependent characteristic potentials (CPs) that relate the variation of the spin-dependent potential $U_{\sigma}$ to voltage and temperature shifts at terminal $\alpha=1,2$.

We treat electron-electron interactions within a mean-field approximation.
The self-consistent determination of $U$ can thus be achieved by solving the Poisson equation $\nabla^{2}\Delta U=-4\pi q$, with $\Delta U=U-U_{\text{eq}}=\sum_{\sigma}\Delta U_{\sigma}$ and
\begin{equation}\label{q}
q=\sum_{\sigma}q_{\sigma}=e\sum_{\alpha,\sigma}\Big[D_{\alpha}^{p}(\sigma)eV_{\alpha}+D_{\alpha}^{e}(\sigma)\theta_{\alpha}\Big]
	+e^{2}\sum_{\sigma}\Pi_{\sigma}\Delta U_{\sigma}\,.
\end{equation}
The charge pileup $q$ is given by the sum of the bare injected charge determined from the spin-dependent particle\cite{but93,chr96} ($p$) and entropic\cite{san13} ($e$) injectivities,
$D_{\alpha}^{p,e}(\sigma)=-\int dE\nu_{\alpha}^{p,e}(E,\sigma)\partial_{E}f$, where $\nu_{\alpha}^{p}(E,\sigma)=(2\pi i)^{-1}\sum_{\beta}\text{Tr}\big[s_{\beta\alpha}^{\dag}\frac{ds_{\beta\alpha}}{dE}\big]$ and $\nu_{\alpha}^{e}(E,\sigma)=(2\pi i)^{-1}\sum_{\beta}\text{Tr}\big[\frac{E-E_{F}}{T}s_{\beta\alpha}^{\dag}\frac{ds_{\beta\alpha}}{dE}\big]$,
and the screening charge $e^{2}\sum_{\sigma}\Pi_{\sigma}\Delta U_{\sigma}$, where $\Pi_{\sigma}$ is the spin-dependent Lindhard function which in the long wavelength limit becomes $\Pi_{\sigma}=\int dE \nu_{\sigma}(E)\partial_{E}f$, with
$\nu_\sigma(E)=\sum_\alpha \nu_{\alpha}^{p}(E,\sigma)$ the spin-$\sigma$ electron density of states.
Then,  the integrated density of states is $D_{\sigma}=\sum_\alpha D_{\alpha}^{p}$. 
	Note, however, that possible $\sigma$ dependences of $D_{\alpha}^{p,e}(\sigma)$ and $\Pi_{\sigma}$ would only appear
in our model for unequal spin populations arising, e.g., from ferromagnetic contacts.
	Thus, for normal metallic contacts the only spin-dependent term in Eq.~\eqref{q} is the screening $\Delta U_{\sigma}$ giving rise to a spin imbalance inside the system.

In the general case, the potential $U(\vec{r})$ is a space-dependent function.
For a practical calculation, we discretize the conductor into the regions illustrated in Fig.~\ref{fig:1}:
$\Omega_{is}$, with $i=1,2$  for the upper and lower edges, $s=\pm$ denoting the helicity,
and dot region with spin $\sigma$.
	The edge states are tunnel-coupled to the dot via hybridization widths $\Gamma_{1s}$ and $\Gamma_{2s}$, which explicitly depend on the helicity $s=\pm$ corresponding to spin channels $\up$($+$) and $\down$($-$).
	The dot is described with a quasilocalized level whose energy $E_{d}$ is controllable by a top gate potential.
	In the wide-band limit, scattering with the dot is well described using a Breit-Wigner form. Hence,
	the reflection probability off the dot is given by $r_{\sigma}=1-t_{\sigma}=\Gamma_{1s}\Gamma_{2s}/|\Lambda_{s}|^{2}$, where $\Lambda_{s}=E_{F}-E_{d}+i\Gamma_{s}/2$ with $\Gamma_{s}=\Gamma_{1s}+\Gamma_{2s}$,
	where $t_{\sigma}$ is the transmission probability.
	Importantly, the helicity $s$-dependence of $\Gamma_{i s}$ ($i=1,2$) disappears for normal contacts, since in this case there is no spin imbalance inside the edge states.
	This leads to spin-independent transmissions $t_{\up}=t_{\down}$ via antidot scattering. As a consequence,
	the linear conductance coefficients are spin-independent and the spin-polarization arises \textit{only} in the nonlinear regime of transport.
		
	The potential $U_{is}$ in each region is assumed to be spatially homogeneous. We describe the Coulomb interaction between the edge states and the dot with a capacitance matrix $C_{is,d\sigma}$.\cite{but93}
	This discrete local potential model captures the essential physics.\cite{chr96,san04}
	The region-specific CPs are then given by $u_{i\alpha}^{\sigma}=(\partial U_{i}^{\sigma}/\partial V_{\alpha})_{\text{eq}}$ and $z_{i\alpha}^{\sigma}=(\partial U_{i}^{\sigma}/\partial\theta_{\alpha})_{\text{eq}}$, and the net charge response for each region can be related to the capacitance matrix via
\begin{multline}\label{Poisson}
q_{is}=e\sum_{\alpha}(D_{is,\alpha}^{p}eV_{\alpha}+D_{is,\alpha}^{e}\theta_{\alpha})+e^{2}\Pi_{is}\Delta U_{is}\\
		=\sum_{\sigma}C_{is,d\sigma}(\Delta U_{is}-\Delta U_{d\sigma}).
\end{multline}
	By solving this, one can determine the potential $U_{i\sigma}=U_{is}$ as a function of the applied voltages and the thermal gradients and obtain the spin-dependent CPs according to Eq.~\eqref{eq:U} for each spin.
	It should be noted that the charge with spin $\sigma=\up$($\down$) in the antidot region is supplied from the edge states with helicity $s=+$($-$) via tunnel coupling since we  neglect spin-flip processes in order to maximize spin-polarization effects.
	For definiteness, we assume that the density of states for all regions are equal, i.e., $D_{is}=D_{s}\equiv D/2$, and the injectivities from the two terminals are symmetric, which amount to $D_{is,\alpha}^{p,e}=D_{s}^{p,e}\equiv D^{p,e}/2$ and $\Pi_{is}=\Pi_{s}\equiv\Pi/2$.

	We consider the case where the conductor is electrically symmetric, i.e., $C_{is,d\sigma}=C_{is}=C_{s}=C/2$ with $C=C_{+}+C_{-}$, but asymmetric in the scattering properties such that $\Gamma_{1s}=(1+\eta)\Gamma/4$ and $\Gamma_{2s}=(1-\eta)\Gamma/4$ with $\Gamma=\Gamma_{+}+\Gamma_{-}$ ($\Gamma_{s}=\Gamma_{1s}+\Gamma_{2s}=\Gamma/2$).
	Experimentally, this would be the general situation for dots closer to one of the edge states. Another possibility
	is to tune the width and the height of the tunnel barriers formed between the resonance and the propagating channels.
	Thus, the coupling asymmetry is described with a nonzero $\eta=(\Gamma_{1}-\Gamma_{2})/\Gamma$ where $\Gamma_{i}=\sum_{s}\Gamma_{is}$.
	From Eqs.~\eqref{eq:U} and \eqref{Poisson}, we find the dot potential
\begin{equation}
\Delta U_{d\sigma}=u_{1\sigma}V_{1}+u_{2\sigma}V_{2}+z_{1\sigma}\theta_{1}+z_{2\sigma}\theta_{2},
\end{equation}	
	 with the corresponding CPs
\begin{align}
&u_{1\up}=u_{2\down}=\frac{1}{2}+\eta c_{\text{sc}},\quad
u_{1\down}=u_{2\up}=\frac{1}{2}-\eta c_{\text{sc}},\label{CPu}\\
&z_{1\up}=z_{2\down}=\frac{D^{e}}{eD^{p}}u_{1\up},\quad
z_{1\down}=z_{2\up}=\frac{D^{e}}{eD^{p}}u_{1\down},\label{CPz}
\end{align}
where $c_{\text{sc}}=[2-2C/e^{2}\Pi]^{-1}=C_\mu/2C$
with $1/C_\mu=1/C+1/e^2D$ the electrochemical capacitance.
Importantly, the CPs become {\em spin-dependent}
(e.g., $u_{1\up}-u_{1\down}=2\eta c_{\text{sc}}$)
whenever $\eta\neq 0$. As a result, we expect electronic
transport to be spin polarized for asymmetric couplings.
Interestingly, the strength of the CPs polarization is determined by the
ratio $C_\mu/C$, similarly to the interaction induced magnetic field asymmetry
in nonlinear mesoscopic transport.\cite{but05} In other words, our effect
has a pure interaction origin and vanishes in the noninteracting limit
($C\to\infty$).

The spin dependence of the nonequilibrium potential response can be easily
understood in the following way. Suppose that the left voltage is lifted with
an amount $\Delta V$ while the right voltage remains unchanged. Then,
both the upper edge with $s=+$ and the lower edge state with $s=-$
carry more charge than their counterparts. Since the dot is, say, more coupled
to the upper edge than to the lower one, effectively more electrons with
spin $\uparrow$ are injected into the dot than electrons with spin $\downarrow$.
We emphasize that this effect will be visible in the nonlinear regime of transport
only since the linear response coefficients are independent of the CPs in Eqs.~\eqref{CPu}
and~\eqref{CPz}.

\section{Weakly nonlinear transport}\label{sec:weakly}
	In order to illustrate the mechanism of spin polarization for the currents, we firstly focus on the weakly nonlinear regime of transport and expand the electronic and heat currents in Eqs.~\eqref{I_sigma} and \eqref{J_sigma} around the equilibrium state, $\mu_{\alpha}=E_{F}$ and $T_{\alpha}=T$, up to second order in the driving fields, $V_{\alpha}$ and $\theta_{\alpha}$:\cite{san13,mea13,lop13}
\begin{multline}
I_{\alpha}^{\sigma}=\sum_{\beta}\Big(G_{\alpha\beta}^{\sigma}V_{\beta}+L_{\alpha\beta}^{\sigma}\theta_{\beta}\Big)\\
	+\sum_{\beta\gamma}\Big(G_{\alpha\beta\gamma}^{\sigma}V_{\beta}V_{\gamma}
	+L_{\alpha\beta\gamma}^{\sigma}\theta_{\beta}\theta_{\gamma}
	+2M_{\alpha\beta\gamma}^{\sigma}V_{\beta}\theta_{\gamma}\Big),\label{elec}
\end{multline}
\begin{multline}
{\cal J}_{\alpha}^{\sigma}=\sum_{\beta}\Big(R_{\alpha\beta}^{\sigma}V_{\beta}+K_{\alpha\beta}^{\sigma}\theta_{\beta}\Big)\\
	+\sum_{\beta\gamma}\Big(R_{\alpha\beta\gamma}^{\sigma}V_{\beta}V_{\gamma}
	+K_{\alpha\beta\gamma}^{\sigma}\theta_{\beta}\theta_{\gamma}+2H_{\alpha\beta\gamma}^{\sigma}V_{\beta}\theta_{\gamma}\Big).\label{heat}
\end{multline}
	These general multi-terminal expressions can easily be applied to our two-terminal setup.
	In Appendix~\ref{appen:A}, we explicitly write down compact expressions using a Sommerfeld expansion
	for illustrative purposes, even though this expansion is valid for low temperatures only.
	Below, we shall numerically evaluate the currents by directly integrating Eqs.~\eqref{I_sigma} and \eqref{J_sigma}
	and compare with the analytic results.
	
	Controlled edge backscattering across the dot is given by the transmission probability
	$t(E_{F})=16(E_{F}-E_{d})^{2}/[16(E_{F}-E_{d})^{2}+\Gamma^{2}]$, which is a spin-independent function
	since $\Gamma_{1s}=\Gamma_{1}/2$, $\Gamma_{2s}=\Gamma_{2}/2$. Hence,
  all linear responses are also spin-independent, i.e., $G_{\alpha\beta}^{\up}=G_{\alpha\beta}^{\down}$, $L_{\alpha\beta}^{\up}=L_{\alpha\beta}^{\down}$, $R_{\alpha\beta}^{\up}=R_{\alpha\beta}^{\down}$, and $K_{\alpha\beta}^{\up}=K_{\alpha\beta}^{\down}$ ($\alpha,\beta=1,2$), as should be [see Eqs.~\eqref{appen:linearG}, \eqref{appen:linearL}, \eqref{appen:linearR}, and \eqref{appen:linearK}]. This is a straightforward consequence
  of the fact that linear coefficients are independent of the screening potential.
	Therefore, spin polarization effects arise in the nonlinear regime of transport only, since nonlinear responses are functions of
	the CPs and these can exhibit spin asymmetries, e.g., $G_{111}^{\up}\ne G_{111}^{\down}$ with a nonzero $\eta$.
	This is clear when we substitute Eq.~\eqref{CPu} into Eq.~\eqref{appen:G111}.

	Hence, in the presence of both voltage and thermal biases with $V_{1}=V$, $V_{2}=0$, $\theta_{1}=\theta$, and $\theta_{2}=0$, the spin-polarized electronic and heat currents read
\begin{multline}\label{eq:Isp}
I_{s}=\big[G_{111}^{\up}-G_{111}^{\down}\big]V^{2}
+\big[L_{111}^{\up}-L_{111}^{\down}\big]\theta^{2}\\
+2\big[M_{111}^{\up}-M_{111}^{\down}\big]V\theta,
\end{multline}
\begin{multline}\label{eq:Jsp}
{\cal J}_{s}=\big[R_{111}^{\up}-R_{111}^{\down}\big]V^{2}
+\big[K_{111}^{\up}-K_{111}^{\down}\big]\theta^{2}\\
+2\big[H_{111}^{\up}-H_{111}^{\down}\big]V\theta.
\end{multline}
	We emphasize that the effects discussed in this work remain the same even if we consider different types of bias configurations such as $V_{1}=V/2$, $V_{2}=-V/2$, $\theta_{1}=-\theta/2$, $\theta_{2}=\theta/2$, which, however, only complicate the algebra within our context. 

	The ordinary charge and heat currents are written by
\begin{multline}\label{eq:Icp}
I_{c}=\big[G_{11}^{\up}+G_{11}^{\down}\big]V+\big[L_{11}^{\up}+L_{11}^{\down}\big]\theta
+\big[G_{111}^{\up}+G_{111}^{\down}\big]V^{2}\\
+\big[L_{111}^{\up}+L_{111}^{\down}\big]\theta^{2}
+2\big[M_{111}^{\up}+M_{111}^{\down}\big]V\theta,
\end{multline}
\begin{multline}\label{eq:Jcp}
{\cal J}_{c}=\big[R_{11}^{\up}+R_{11}^{\down}\big]V+\big[K_{11}^{\up}+K_{11}^{\down}\big]\theta
+\big[R_{111}^{\up}+R_{111}^{\down}\big]V^{2}\\
+\big[K_{111}^{\up}+K_{111}^{\down}\big]\theta^{2}
+2\big[H_{111}^{\up}+H_{111}^{\down}\big]V\theta.
\end{multline}

	Applying the relevant nonlinear coefficients in Appendix~\ref{appen:A} to Eqs.~\eqref{eq:Isp} and \eqref{eq:Jsp},
	we find
\begin{multline}\label{eq:IsNor}
I_{s}=-\frac{e^{3}}{h}(u_{1\up}-u_{1\down})t'V^{2}
	-\frac{e^{2}\pi^{2}k_{B}^{2}T}{3h}(z_{1\up}-z_{1\down})t''\theta^{2}\\
-\frac{e^{3}}{h}\bigg[\frac{\pi^{2}k_{B}^{2}T}{3e}(u_{1\up}-u_{1\down})t''+(z_{1\up}-z_{1\down})t'\bigg]V\theta,
\end{multline}
\begin{multline}\label{eq:JsNor}
\calj_{s}=-\frac{e^{2}\pi^{2}(k_{B}T)^{2}}{3h}(u_{1\up}-u_{1\down})t''V^{2}
-\frac{e\pi^{2}k_{B}^{2}T}{3h}(z_{1\up}-z_{1\down})t'\theta^{2}\\
-\frac{e^{2}\pi^{2}(k_{B}T)^{2}}{3h}\bigg[\frac{1}{eT}(u_{1\up}-u_{1\down})t'+(z_{1\up}-z_{1\down})t''\bigg]V\theta,
\end{multline}
where $t\equiv t(E_{F})$, $t'\equiv\partial_{E}t(E)|_{E=E_{F}}$, and $t''\equiv\partial^{2}_{E}t(E)|_{E=E_{F}}$.
	These expressions are central to our results.
	The spin-polarized electronic and heat currents indeed appear when the potential response via antidot scattering is different with respect to each spin component, i.e., either $u_{1\up}-u_{1\down}\ne0$ or $z_{1\up}-z_{1\down}\ne0$.
	Using the CPs in Eqs.~\eqref{CPu} and \eqref{CPz} explicitly, one can finally write
\begin{multline}\label{eq:Is_eta}
I_{s}=-\eta c_{\text{sc}}\bigg(\frac{2e^{3}}{h}t'V^{2}
	+\frac{2e\pi^{2}k_{B}^{2}T}{3h}\frac{D^{e}}{D^{p}}t''\theta^{2}\\
+\frac{2e^{2}}{h}\bigg[\frac{\pi^{2}k_{B}^{2}T}{3}t''
+\frac{D^{e}}{D^{p}}t'\bigg]V\theta\bigg),
\end{multline}
\begin{multline}\label{eq:Js_eta}
\calj_{s}=-\eta c_{\text{sc}}\bigg(\frac{2e^{2}\pi^{2}(k_{B}T)^{2}}{3h}t''V^{2}
+\frac{2\pi^{2}k_{B}^{2}T}{3h}\frac{D^{e}}{D^{p}}t'\theta^{2}\\
+\frac{2e\pi^{2}(k_{B}T)^{2}}{3h}\bigg[\frac{1}{T}t'+\frac{D^{e}}{D^{p}}t''\bigg]V\theta\bigg).
\end{multline}
	Note that the spin-polarization of both currents is directly proportional to the asymmetry parameter $\eta$
	and the interaction parameter $c_{\text{sc}}$. Hence, the asymmetrically coupled quantum antidot plays the role of a spin filter.
	In contrast, as shown in Eqs.~\eqref{eq:Icp} and \eqref{eq:Jcp}, the effect of the potential response on the usual electronic and heat currents can be represented by the sum $u_{1\up}+u_{1\down}$ and $z_{1\up}+z_{1\down}$ rather than the difference.
	Due to helicity, we have $u_{1\up}+u_{1\down}=1$ and $z_{1\up}+z_{1\down}=D^{e}/eD^{p}$ from Eqs.~\eqref{CPu} and \eqref{CPz}, independently of the asymmetry:
\begin{multline}\label{eq:Ic_eta}
I_{c}=\frac{2e^{2}}{h}tV
+\frac{2e\pi^{2}k_{B}^{2}T}{3h}t'\theta
		+\frac{e\pi^{2}k_{B}^{2}}{3h}\bigg(t'-T\frac{D^{e}}{D^{p}}t''\bigg)\theta^{2}\\
+\frac{e^{2}}{h}\bigg(\frac{\pi^{2}k_{B}^{2}T}{3}t''-\frac{D^{e}}{D^{p}}t'\bigg)V\theta,
\end{multline}
\begin{multline}\label{eq:Jc_eta}
\calj_{c}=\frac{2e\pi^{2}(k_{B}T)^{2}}{3h}t'V
+\frac{2\pi^{2}k_{B}^{2}T}{3h}t\theta
	-\frac{e^{2}}{h}\bigg(t+\frac{\pi^{2}(k_{B}T)^{2}}{6}t''\bigg)V^{2}\\
+\frac{\pi^{2}k_{B}^{2}}{3h}\bigg(t-T\frac{D^{e}}{D^{p}}t'\bigg)\theta^{2}
+\frac{e\pi^{2}k_{B}^{2}T}{3h}\bigg(t'-T\frac{D^{e}}{D^{p}}t''\bigg)V\theta.
\end{multline}
	Remarkably, the second-order electric response $G_{111}^{\up}+G_{111}^{\down}$ cancels out because this term contains the screening effect with a factor $1-(u_{1\up}+u_{1\down})$ [Eq.~\eqref{appen:G111}], which is always zero for normal contacts due to helical nature of the edge states.
	It should be emphasized that this cancellation is not originated from our specific bias setup $V_{1}=V,~V_{2}=0$.
	Indeed, even for a general voltage bias configuration, i.e., $V_{1}=\xi V$ and $V_{2}=(\xi-1)V$ with $0\le\xi\le1$, the second order effect of voltage driving can be written as $\sum_{\sigma}G_{111}^{\sigma}(V_{1}-V_{2})^{2}=\sum_{\sigma}G_{122}^{\sigma}(V_{1}-V_{2})^{2}=-\sum_{\sigma}G_{211}^{\sigma}(V_{1}-V_{2})^{2}=0$, due to gauge invariance and current conservation.
	Therefore, the charge current in the isothermal case, i.e., $\theta_{1}=\theta_{2}=0$, is always given by $I_{c}=(2e^{2}/h)tV$ up to order $V^3$. This absence of rectification effects in our two-dimensional topological insulator system is in stark contrast
	with small conductors coupled to normal reservoirs, in which the $V^{2}$ term is generally present.\cite{son98, lin00, sho01, fle02, but03, gon04, hac04, seg05}

\section{Numerical results}\label{Numerical}

	In the previous section, we discussed the underlying spin-filter mechanism in an intuitive way,
	deriving expressions valid in the weakly nonlinear regime, as shown in Eqs.~\eqref{eq:IsNor} and \eqref{eq:JsNor}.
	These analytic results are also based on a Sommerfeld expansion, which is appropriate at low temperatures.
	To extend the validity of our conclusions for both strong nonlinearities and high temperatures, we now evaluate the currents numerically via direct integration of Eqs.~\eqref{I_sigma} and \eqref{J_sigma} without any further assumption. Our only limitation is the mean-field approximation, thus neglecting strong electron-electron correlations in our system. Below, we discuss the isothermal ($\theta_{1}=\theta_{2}=0$) and isoelectric ($V_{1}=V_{2}=0$) cases separately. Finally, we consider the general case ($V_{1}=V, \theta_{1}=\theta$) for which, interestingly, pure spin currents can be generated.

\begin{figure}[t]
  \centering
\includegraphics[angle=270,width=0.45\textwidth,clip]{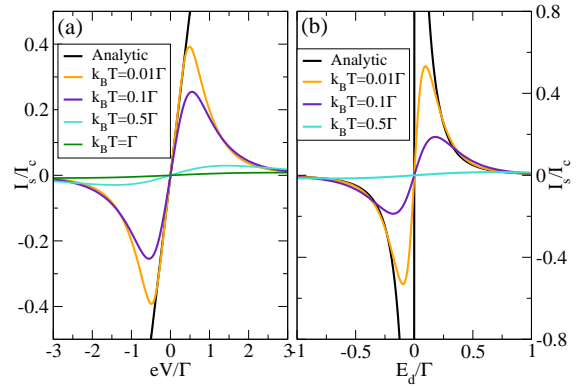}
\caption{(Color online) Plots of $I_s/I_c$ versus (a) voltage bias $eV/\Gamma$ at $E_{d}/\Gamma=0.25$ and (b) antidot level $E_{d}/\Gamma$ at $eV/\Gamma=0.25$, for several background temperatures $k_{B}T$ in the isothermal case.
In all cases, we use $\eta=c_{\text{sc}}=0.5$ and $E_{F}=0$.
}\label{fig2}
\end{figure}
\begin{figure}[t]
  \centering
\includegraphics[angle=270,width=0.45\textwidth, clip]{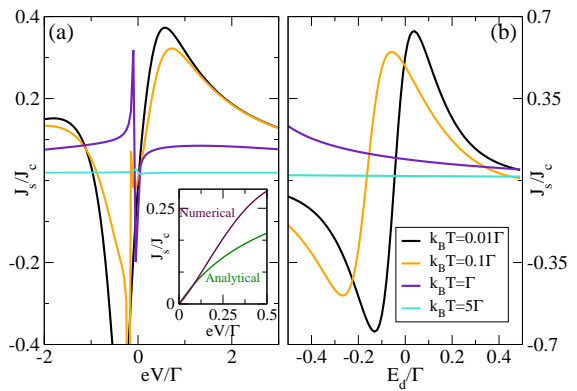}\\
\caption{(Color online) Plots of $\calj_s/\calj_c$ versus (a) voltage bias $eV/\Gamma$ at $E_{d}/\Gamma=0.2$ and (b) antidot level $E_{d}/\Gamma$ at $eV/\Gamma=0.25$, for several background temperatures $k_{B}T$ in the isothermal case.
In the inset of (a), an analytic result is shown in comparison with the numerical one at $k_{B}T/\Gamma=0.1$.
Parameters used are $\eta=c_{\text{sc}}=0.5$. 
Note that since at moderate voltages the Joule heating present in $\mathcal{J}_c$ dominates the spin heat flow
quickly becomes a nonlinear function of $V$.}\label{fig3}
\end{figure}

\subsection{Voltage-driven transport: isothermal case}

	In Fig.~\ref{fig2}(a), we plot the dimensionless ratio $I_s/I_c$ between the spin-polarized current and the charge flux
	as a function of the voltage bias $V$ for a given antidot level position $E_d$.
	At low voltages, we observe a linear dependence of $I_s/I_c$ with $V$, in agreement with the analytical results.
	We note that for $\theta_{1}=\theta_{2}=0=V_{2}$ and $V_{1}=V$, the spin-polarized current in Eq.~\eqref{eq:Is_eta}
	reduces to
	\begin{equation}\label{eq_isiso}
	I_{s}=-\frac{2e^{3}}{h}\eta c_{\text{sc}}t'V^{2}\,,
	\end{equation}
	while the charge current is simply given by $I_{c}=(2e^{2}/h)tV$, both to leading order in a voltage expansion
	for low $T$.
	Therefore, the degree of polarization $I_s/I_c$ increases with voltage for small $V$. At higher voltages, the polarization
	decreases when $V$ is larger than $\Gamma/e$ because charge fluctuations are quenched. In Fig.~\ref{fig2}(b),
	we show the gate tuning of $I_s/I_c$, which is depicted for a fixed bias. Again, the maximal polarization is attained
	when the dot level is above or below the Fermi energy on the scale of the hybridization width $\Gamma$
	because Eq.~\eqref{eq_isiso} shows that the spin current is proportional to $t'$, which is a function with an energy
	dependence governed by $\Gamma$ in the Breit-Wigner approximation.
	Furthermore, our results show that the polarization decreases when the background temperature $T$ increases
	since large temperatures tend to smear out the energy dependence of the scattering matrix, an essential ingredient
	of our spin-filter effect.
	
	Figure~\ref{fig3}(a) displays the spin polarization of the heat current, defined as $\calj_s/\calj_c$, as a function of the
	bias voltage. For small $V$ in the isothermal case, Eq.~\eqref{eq:Js_eta} yields
	\begin{equation}
		\calj_{s}=-\eta c_{\text{sc}}(2e^{2}\pi^{2}/3h)(k_{B}T)^{2}t''V^{2}\,.
	\end{equation}
	This can be seen as the leading-order spin-polarized\cite{gra06} nonlinear Peltier effect.\cite{kul94, bog99, zeb07}
	In turn, the heat flux associated to charge transport is given, to lowest order in $V$, by $\calj_{c}=(2e\pi^{2}/3h)(k_{B}T)^{2}t'V-(e^{2}/h)[t+(\pi^{2}/6)(k_{B}T)^{2}t'']V^{2}$
	[we set $\theta_{1}=\theta_{2}=0=V_{2}$ and $V_{1}=V$ in Eq.~\eqref{eq:Jc_eta}], where the conventional Peltier
	coefficient and the Joule heating term are clearly shown. Since the latter dominates even at low $V$, the spin
	polarization quickly departs from the linear dependence, see the inset of Fig.~\ref{fig3}(a). Moreover, we observe
	an asymmetry between positive and negative voltages due to the heat current being, in general, asymmetric
	with respect to energy integration due to the $\mu=E_F+eV$ term in Eq.~\eqref{J_sigma}. Recent experiments
	with scanning tunneling microscope probes coupled to molecules attached to substrate precisely observe an asymmetric
	heat dissipation in the charge sector.\cite{lee13} Here, we predict that the same phenomenon will occur for the spin degree
	of freedom and that it can be manipulated either changing the base temperature or the dot level position,
	see Fig.~\ref{fig3}(b).

\subsection{Temperature-driven transport: isoelectric  case}
\begin{figure}[t]
  \centering
\includegraphics[angle=270,width=0.5\textwidth, clip]{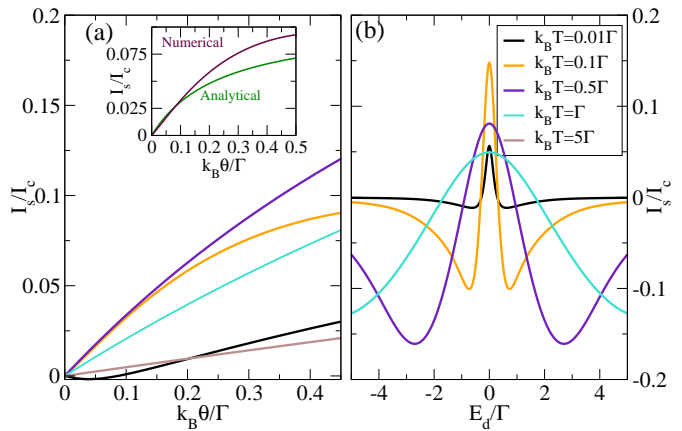}\\
\caption{(Color online) Plots of $I_s/I_c$ versus (a) thermal gradients $k_{B}\theta/\Gamma$ at $E_{d}/\Gamma=0.2$ and (b) antidot level $E_{d}/\Gamma$ at $k_{B}\theta/\Gamma=0.25$, for several background temperatures $k_{B}T$ in the isoelectric case.
In the inset of (a), an analytic result is shown in comparison with the numerical one at $k_{B}T/\Gamma=0.1$.
We use $\eta=c_{\text{sc}}=0.5$ and $E_{F}=0$.
}\label{fig4}
\end{figure}

	We now consider the case of an applied temperature bias such as $\theta_{1}=\theta$ and $\theta_{2}=0$
	for equal electrochemical potentials $V_{1}=V_{2}=0$. To leading order in a $\theta$ expansion, the spin-dependent
	current becomes at low $T$
	\begin{equation}\label{eq_Isisoel}
		I_{s}=-\eta c_{\text{sc}}(2e\pi^{2}k_{B}^{2}T/3h)(D^{e}/D^{p})t''\theta^{2}\,.
	\end{equation}
	Similarly to the isothermal case [cf. Eq.~\eqref{eq_isiso}], the spin current is purely nonlinear in the driving field.
	Nevertheless, unlike the isothermal case $I_s$ in the isoelectric case depends not only on the particle injectivity but also on the entropic contribution since
	the temperature dependence of the transmission is determined, to leading order, by the carrier energy measured
	with regard to $E_F$.\cite{san13} We also note that $I_s$ vanishes if the background temperature $T$ tends to zero,
	thereby our thermal spin generation has a thermoelectric character like the spin Seebeck effect.\cite{Uchida,jaw10,sla10}
	In fact, the charge current is simply given
	by the thermocurrent expression $I_{c}=(2e\pi^{2}k_{B}^{2}T/3h)t'\theta$ up to $\mathcal{O}(\theta)$.
	Hence, the spin-polarization ratio $I_s/I_c$ is a linear function of $\theta$ at low $\theta$. This is confirmed
	with our numerical results in Fig.~\ref{fig4}(a). In Fig.~\ref{fig4}(b) we show that the spin-filter effect can be,
	to a large extent, tuned with a gate voltage for a fixed value of $\theta$, which can even reverse the sign
	of $I_s/I_c$. In contrast to the isothermal case, the spin polarization degree vanishes
	for very low temperatures except for $E_d$ close to the leads' Fermi energy. It is precisely at this energy
	for which the isoelectric $I_s$ is more sensitive to changes in $\theta$, in agreement with Eq.~\eqref{eq_Isisoel}.
	
\begin{figure}[t]
  \centering
\includegraphics[angle=270,width=0.5\textwidth, clip]{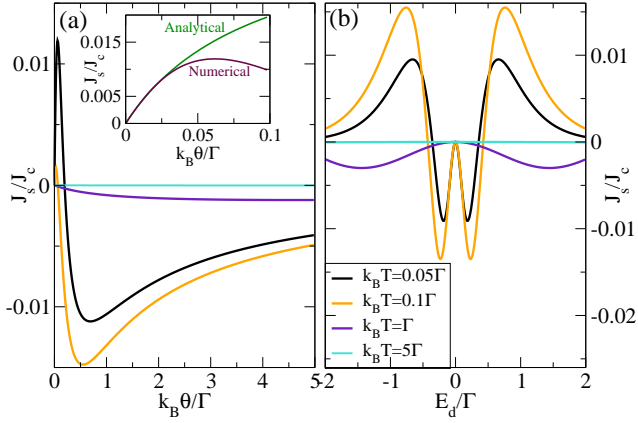}\\
\caption{(Color online) Plots of $\mathcal{J}_s/\mathcal{J}_c$ versus (a) thermal gradients $k_{B}\theta/\Gamma$ at $E_{d}/\Gamma=0.3$ and (b) antidot level $E_{d}/\Gamma$ at $k_{B}\theta/\Gamma=0.25$, for several background temperatures $k_{B}T$ in the isoelectric case.
In the inset of (a), an analytic result is shown in comparison with the numerical one at $k_{B}T/\Gamma=0.05$.
We use $\eta=c_{\text{sc}}=0.5$ and $E_{F}=0$.
}\label{fig5}
\end{figure}


	The heat current can also become spin polarized upon the application of a thermal gradient because the generalized
	thermal conductance depends on the spin index, see Eq.~\eqref{eq:K111}.
	For $\theta_{1}=\theta$ and $V_{1}=V_{2}=0=\theta_{2}$ we find
	\begin{equation}
		\calj_{s}=-\eta c_{\text{sc}}(2\pi^{2}k_{B}^{2}T/3h)(D^{e}/D^{p})t'\theta^{2}
	\end{equation}
	to leading order in the temperature bias. The heat current due to charge transport is given by
	$\calj_{c}=(2\pi^{2}k_{B}^{2}T/3h)t\theta + \mathcal{O}(\theta)^2$ at low $T$. Therefore, the ratio
	$\calj_{s}/\calj_{c}$ is generally nonzero for increasing $\theta$, see Fig.~\ref{fig5}(a).
	Interestingly, at resonance ($E_d=E_F$) the spin polarization of the heat current becomes zero
	[Fig.~\ref{fig5}(b)] while the electric current counterpart shows a local maximum [Fig.~\ref{fig4}(b)],
	indicating that the spin-filter mechanism of a QSH antidot acts differently to electric and heat currents.

\subsection{Thermoelectric transport: pure spin currents}\label{Seebeck}
\begin{figure}[t]
  \centering
\includegraphics[angle=270,width=0.5\textwidth, clip]{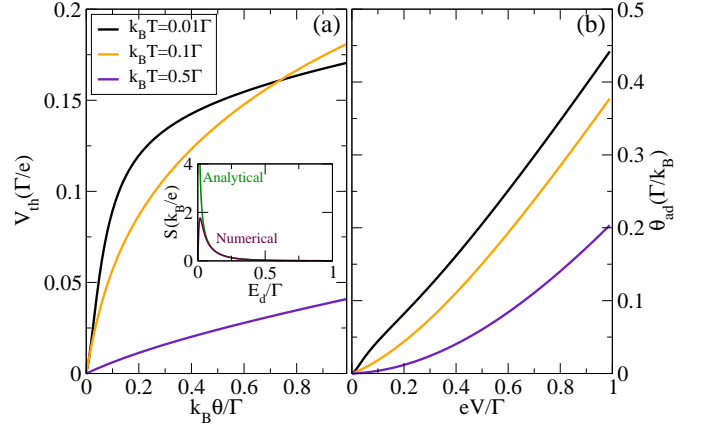}\\
\caption{(Color online) Plots of generated (a) thermovoltage $V_{\text{th}}$ versus applied thermal gradient $k_{B}\theta/\Gamma$ and (b) adiabatic thermal gradient $\theta_{\text{ad}}$ versus voltage bias $eV/\Gamma$, at $E_{d}=0.1\Gamma$ for several background temperatures $k_{B}T$.
In the inset of (a), the Seebeck coefficient with analytic and numerical results at $k_{B}T=0.01\Gamma$ are shown as a function of resonance level $E_d/\Gamma$.
Parameters are $\eta=c_{\text{sc}}=0.5$ and $E_{F}=0$.
}\label{fig6}
\end{figure}

	We have shown above that thermal gradients can generate  spin-polarized thermocurrents $I_{s}\neq 0$,
	as a synergistic combination of thermoelectric and spintronic effects.\cite{Uchida,jaw10,sla10}
	We now prove that it is even possible to create pure spin currents, i.e., $I_{s}\neq 0$ for vanishingly small
	charge current, $I_{c}= 0$. The latter condition can be easily achieved in open-circuit conditions, in which
	case a thermovoltage $V_\text{th}$ is generated in response to a temperature bias $\theta$.
	In Fig.~\ref{fig6}(a) we plot the numerically calculated set of biases $\{\theta,V\}$ which
	satisfy the expression $I_{c}(V_\text{th},\theta)= 0$ as a function of $\theta$. As expected,
	at low temperature bias the thermovoltage shows a linear dependence because the Seebeck coefficient,
	$S=V_\text{th}/\theta$, is constant for small thermal gradients. With increasing $\theta$,
	the thermovoltage acquires a nonlinear component.\cite{san13,fah13}

\begin{figure}[t]
  \centering
\includegraphics[angle=270,width=0.5\textwidth, clip]{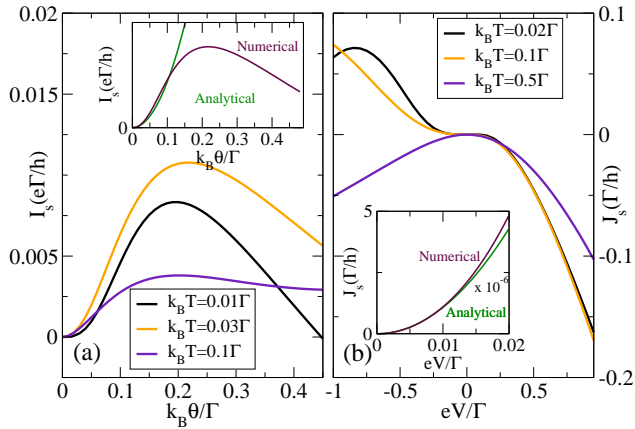}\\
\caption{(Color online) Plots of (a) pure spin currents $I_s$ versus thermal gradient $k_{B}\theta/\Gamma$ and (b) pure spin heat currents $\calj_s$ versus voltage bias $eV/\Gamma$, at $E_{d}/\Gamma=0.25$ for several background temperatures $k_{B}T$.
The insets compare the analytic and numerical results at (a) $k_{B}T=0.03\Gamma$ and (b) $k_{B}T=0.02\Gamma$, where the latter comparision has been made in a very small bias range where $\calj_{s}$ is positive.
We use $\eta=c_{\text{sc}}=0.5$ and $E_{F}=0$.
}\label{fig7}
\end{figure}

	Substituting $V$ with $V_{\rm th}(\theta)$ in the expression for $I_s$ we find the pure spin current
\begin{multline}\label{IsPure}
I_{s}=\eta c_{\text{sc}}\frac{2e\pi^{2}k_{B}^{2}T}{3h}\Bigg(\frac{\pi^{2}k_{B}^{2}T}{3}\bigg[\frac{t't''}{t}-\frac{(t')^{3}}{t^{2}}\bigg]\\
	+\frac{D^{e}}{D^{p}}\bigg[\frac{(t')^{2}}{t}-t''\bigg]\Bigg)\theta^{2},
\end{multline}
up to leading order in $\theta$. Figure~\ref{fig7}(a) shows the numerical results for pure $I_s$ beyond the quadratic
regime (the inset displays a comparison with the analytical results). We observe that  the amplitude of $I_s$ firstly increases as $T$
is enhanced (here, it is shown from $k_BT/\Gamma=0.01$ to $k_BT/\Gamma=0.03$) and then decreases (from $k_BT/\Gamma=0.03$ to $k_BT/\Gamma=0.1$), exhibiting a nonmonotonic behavior with $T$.

Our device also creates \textit{pure spin heat flows} using electric means only. We first solve the equation
$\calj_c(V,\theta_\text{ad})=0$, which amounts to adiabatically isolating the sample. This yields
a generated thermal bias $\theta_\text{ad}$ in response to the applied voltage $V$,
see Fig.~\ref{fig6}(b). $\theta_\text{ad}$ is an increasing function of $V$ since a positive thermal gradient
compensates the current flowing through the system. The effect is less pronounced for higher background
temperatures $T$ because more electrons become thermally excited for increasing $T$. We then
substitute $\theta_\text{ad}(V)$ in the $\calj_{s}$ expression  and find,
\begin{multline}\label{JsPure}
\calj_{s}=\eta c_{\text{sc}}\frac{2e^{2}\pi^{2}(k_{B}T)^{2}}{3h}\Bigg(\bigg[\frac{(t')^{2}}{t}-t''\bigg]\\
+T\frac{D^{e}}{D^{p}}\bigg[\frac{t't''}{t}-\frac{(t')^{3}}{t^{2}}\bigg]\Bigg)V^{2}.
\end{multline}
up to leading order in $V$. We plot in Fig.~\ref{fig7}(b) the pure spin heat current $\calj_s$
as a function of the bias voltage. At low $V$, our numerical results agree with Eq.~\eqref{JsPure} (see the inset).
For higher voltage, the results are also in qualitative agreement with $I_s$ because $|\calj_s|$ increases to higher values of $T$ (here it is shown up to $k_BT/\Gamma=0.1$), beyond which the amplitude of $\calj_s$ starts to decrease.

\section{Conclusions}\label{sec:Con}
Two-dimensional topological insulators with controlled backscattering
present a rich spin dynamics which can be manipulated with external
gate potentials and background temperatures. We have demonstrated
that spin-polarized currents can be generated in a two-terminal
quantum spin Hall systems coupled to normal contacts.
Neither Zeeman fields nor ferromagnetic materials are needed
in the implementation of our effect. The spin dependence is purely
induced by interactions and arises in the nonequilibrium screening
potential of the conductor in the response to either voltage
or temperature shifts applied to the contacts. Importantly,
pure spin currents can be created using the Seebeck effect.
The spin-polarization mechanism also works for the heat current,
in which case a pure spin heat flow is generated for adiabatically
isolated samples. 

Our discussion ignores spin-flip processes and Coulomb blockade effects.
The former will be detrimental to our spin filtering operational principle
if spin-flip transitions preserve the momentum.\cite{Ste14}
The latter will have a less clear effect.
Our theory shows that the screening potential becomes spin-independent in the noninteracting limit, i.e., $C\to\infty$ in Eqs.~\eqref{CPu} and \eqref{CPz}. The spin-filtering effect becomes stronger as $C\to0$. Therefore, strong interaction would favor the generation of spin currents and single charge effects are expected to maintain the effects discovered in our work.
However, if Coulomb blockade allows the spin-flip transitions, a more careful analysis should be performed.
Spin-increasing and spin-decreasing
transitions have been experimentally reported.\cite{pot03}
In addition, the  impact of spin-blockade phenomena\cite{ono02} deserves further investigation.

In general, there is considerable scope to extend our model and treat
different situations. For instance, one could consider the competition between
the spin-polarization effects discussed here and spin filtering inherent
to ferromagnetic contacts or Zeeman splittings. Inclusion of these
influences in our theoretical model would be straightforward. Another interesting
possibility would be the study of the thermodynamic efficiency,
a subject of practical importance that has recently attracted a good deal of attention,
especially in quantum conductors.\cite{ben13} 

\section{Acknowledgments}
This research was supported by MINECO under Grant No. FIS2011-23526,
the Kavli Institute for Theoretical Physics through NSF grant PHY11-25915
and the National Research Foundation of Korea (NRF) grants
funded by the Korea government (MSIP) (No.~2011-0030046).
  
\appendix

\section{Coefficients in Sommerfeld expansion}\label{appen:A}
	In a two-terminal setup ignoring the spin-flip scattering, the current conservation condition gives $A_{11}^{\sigma}=A_{22}^{\sigma}=-A_{12}^{\sigma}=-A_{21}^{\sigma}=t^{\sigma}(E)$ where $t^{\sigma}(E)$ is the spin-dependent transmission probability.
	One can find linear and nonlinear coefficients\cite{san13,lop13} in Eqs.~\eqref{elec} and \eqref{heat} to leading order of the Sommerfeld expansion:
\begin{align}
&G_{11}^{\sigma}=G_{22}^{\sigma}=-G_{12}^{\sigma}=-G_{21}^{\sigma}=\frac{e^{2}}{h}t^{\sigma}(E_{F}),\label{appen:linearG}\\
&L_{11}^{\sigma}=L_{22}^{\sigma}=-L_{12}^{\sigma}=-L_{21}^{\sigma}=\frac{e\pi^{2}k_{B}^{2}T}{3h}\frac{\partial t^{\sigma}(E)}{\partial E}\bigg|_{E_{F}},\label{appen:linearL}\\
&R_{11}^{\sigma}=R_{22}^{\sigma}=-R_{12}^{\sigma}=-R_{21}^{\sigma}=\frac{e\pi^{2}(k_{B}T)^{2}}{3h}\frac{\partial t^{\sigma}(E)}{\partial E}\bigg|_{E_{F}},\label{appen:linearR}\\
&K_{11}^{\sigma}=K_{22}^{\sigma}=-K_{12}^{\sigma}=-K_{21}^{\sigma}=\frac{\pi^{2}k_{B}^{2}T}{3h}t^{\sigma}(E_{F}),\label{appen:linearK}
\end{align}

\begin{subequations}\label{appen:G_non}
\begin{align}
G_{111}^{\sigma}&=\frac{e^{3}}{h}\frac{\partial t^{\sigma}(E)}{\partial E}\bigg|_{E_{F}}\bigg(\frac{1}{2}-u_{1\sigma}\bigg),\label{appen:G111}\\
G_{122}^{\sigma}&=\frac{e^{3}}{h}\frac{\partial t^{\sigma}(E)}{\partial E}\bigg|_{E_{F}}\bigg(u_{2\sigma}-\frac{1}{2}\bigg),\\
G_{211}^{\sigma}&=\frac{e^{3}}{h}\frac{\partial t^{\sigma}(E)}{\partial E}\bigg|_{E_{F}}\bigg(u_{1\sigma}-\frac{1}{2}\bigg),
\end{align}
\end{subequations}

\begin{subequations}
\begin{align}
L_{111}^{\sigma}&=\frac{e\pi^{2}k_{B}^{2}}{6h}\bigg[\frac{\partial t^{\sigma}(E)}{\partial E}
			-2ez_{1\sigma}T\frac{\partial^{2}t^{\sigma}(E)}{\partial E^{2}}\bigg]_{E_{F}},\\
L_{122}^{\sigma}&=-\frac{e\pi^{2}k_{B}^{2}}{6h}\bigg[\frac{\partial t^{\sigma}(E)}{\partial E}
			-2ez_{2\sigma}T\frac{\partial^{2}t^{\sigma}(E)}{\partial E^{2}}\bigg]_{E_{F}},\\
L_{211}^{\sigma}&=-\frac{e\pi^{2}k_{B}^{2}}{6h}\bigg[\frac{\partial t^{\sigma}(E)}{\partial E}
			-2ez_{1\sigma}T\frac{\partial^{2}t^{\sigma}(E)}{\partial E^{2}}\bigg]_{E_{F}},
\end{align}
\end{subequations}

\begin{subequations}
\begin{align}
M_{111}^{\sigma}&=-\frac{e^{3}}{2h}\bigg[\frac{\partial t^{\sigma}(E)}{\partial E}z_{1\sigma}+\frac{\pi^{2}k_{B}^{2}T}{3e}\frac{\partial^{2}t^{\sigma}(E)}{\partial E^{2}}(u_{1\sigma}-1)\bigg]_{E_{F}},\\
M_{121}^{\sigma}&=\frac{e^{3}}{2h}\bigg[\frac{\partial t^{\sigma}(E)}{\partial E}z_{1\sigma}-\frac{\pi^{2}k_{B}^{2}T}{3e}\frac{\partial^{2}t^{\sigma}(E)}{\partial E^{2}}u_{2\sigma}\bigg]_{E_{F}},
\end{align}
\end{subequations}

\begin{subequations}
\begin{align}
R_{111}^{\sigma}&=-\frac{e^{2}}{2h}\bigg[t^{\sigma}(E)+
		\frac{\pi^{2}(k_{B}T)^{2}}{6}\frac{\partial^{2}t^{\sigma}(E)}{\partial E^{2}}(4u_{1\sigma}-1)\bigg]_{E_{F}},\\
R_{122}^{\sigma}&=-\frac{e^{2}}{2h}\bigg[t^{\sigma}(E)+
		\frac{\pi^{2}(k_{B}T)^{2}}{6}\frac{\partial^{2}t^{\sigma}(E)}{\partial E^{2}}(3-4u_{2\sigma})\bigg]_{E_{F}},\\
R_{211}^{\sigma}&=-\frac{e^{2}}{2h}\bigg[t^{\sigma}(E)+
		\frac{\pi^{2}(k_{B}T)^{2}}{6}\frac{\partial^{2}t^{\sigma}(E)}{\partial E^{2}}(3-4u_{1\sigma})\bigg]_{E_{F}},
\end{align}
\end{subequations}

\begin{subequations}\label{eq:K111}
\begin{align}
K_{111}^{\sigma}&=\frac{\pi^{2}k_{B}^{2}}{6h}\bigg[t^{\sigma}(E)-2ez_{1\sigma}T\frac{\partial t^{\sigma}(E)}{\partial E}\bigg]_{E_{F}},\\
K_{122}^{\sigma}&=-\frac{\pi^{2}k_{B}^{2}}{6h}\bigg[t^{\sigma}(E)-2ez_{2\sigma}T\frac{\partial t^{\sigma}(E)}{\partial E}\bigg]_{E_{F}},\\
K_{211}^{\sigma}&=-\frac{\pi^{2}k_{B}^{2}}{6h}\bigg[t^{\sigma}(E)-2ez_{1\sigma}T\frac{\partial t^{\sigma}(E)}{\partial E}\bigg]_{E_{F}},
\end{align}
\end{subequations}

\begin{subequations}
\begin{align}
H_{111}^{\sigma}&=\frac{e^{2}\pi^{2}(k_{B}T)^{2}}{6h}\bigg[\frac{1}{eT}\frac{\partial t^{\sigma}(E)}{\partial E}(1-u_{1\sigma})-\frac{\partial^{2} t^{\sigma}(E)}{\partial E^{2}}z_{1\sigma}\bigg]_{E_{F}},\\
H_{121}^{\sigma}&=\frac{e^{2}\pi^{2}(k_{B}T)^{2}}{6h}\bigg[\frac{1}{eT}\frac{\partial t^{\sigma}(E)}{\partial E}(1-u_{2\sigma})+\frac{\partial^{2} t^{\sigma}(E)}{\partial E^{2}}z_{1\sigma}\bigg]_{E_{F}},
\end{align}
\end{subequations}
where $t^{\sigma}(E_{F})=1-\Gamma_{1s}\Gamma_{2s}/|\Lambda_{s}|^{2}$ with $\Lambda_{s}=E_{F}-E_{d}+i\Gamma_{s}/2$, $\Gamma_{s}=\Gamma_{1s}+\Gamma_{2s}$, and $s=\pm$ corresponding to $\sigma=\up,\down$ interchangeably.

\end{document}